\documentclass[letterpaper, 10 pt, conference]{ieeeconf}
\usepackage{graphics} % for pdf, bitmapped graphics files
\usepackage{epsfig} % for postscript graphics files
\usepackage{mathptmx} % assumes new font selection scheme installed
\usepackage{times} % assumes new font selection scheme installed
\usepackage{amsmath} % assumes amsmath package installed
\usepackage{amssymb}  % assumes amsmath package installed
\usepackage{authblk}

\usepackage{graphicx}
\usepackage{dirtytalk}
\usepackage{url}
\usepackage{multirow}
\usepackage{multicol}
\usepackage{ragged2e}
\usepackage{balance}

\begin{document}

\title{\LARGE \bf
Hierarchical Fine-Tuning for joint Liver Lesion Segmentation \\ and Lesion Classification in CT}

\author[1]{Michal Heker}
\author[1]{Avi Ben-Cohen}
\author[1]{Hayit Greenspan}
\affil[1]{Department of Biomedical Engineering, Faculty of Engineering,Tel-Aviv University, Tel-Aviv, Israel}

\maketitle
\thispagestyle{empty}
\pagestyle{empty}

%%%%%%%%%%%%%%%%%%%%%%%%%%%%%%%%%%%%%%%%%%%%%%%%%%%%%%%%%%%%%%%%%%%%%%%%%%%%%%%%
\begin{abstract}
% Abstract - write it based on your paper text starting with the need, the motivation, the proposed methodology with a brief description, most important results, conclusions.
We present an automatic method for joint liver lesion segmentation and classification using a hierarchical fine-tuning framework. Our dataset is small, containing 332 2-D CT examinations with lesion annotated into 3 lesion types: cysts, hemangiomas, and metastases. Using a cascaded U-net that performs segmentation and classification simultaneously, we trained a strong lesion segmentation model on the dataset of MICCAI 2017 Liver Tumor Segmentation (LiTS) Challenge. We used the trained weights to fine-tune a slightly modified model to obtain improved lesion segmentation and classification, on the smaller dataset. Since pre-training was done with similar data on a related task, we were able to learn more representative features (especially higher-level features in the U-Net's encoder), and improve pixel-wise classification results. We show an improvement of over 10\% in Dice score and classification accuracy, compared to a baseline model. We further improve the classification performance by hierarchically freezing the encoder part of the network and achieve an improvement of over 15\% in Dice score and classification accuracy. We compare our results with an existing method and show an improvement of 14\% in the success rate and 12\% in the classification accuracy.

\end{abstract}

% \begin{IEEEkeywords}
% liver lesions, lesion segmentation, lesion classification, fine-tuning, CT.
% \end{IEEEkeywords}

%%%%%%%%%%%%%%%%%%%%%%%%%%%%%%%%%%%%%%%%%%%%%%%%%%%%%%%%%%%%%%%%%%%%%%%%%%%%%%%%
\section{INTRODUCTION}
% Motivation- field of liver cancer
% Liver cancer is one of the most common sites for metastatic cancer.
According to the American Cancer Society's estimates, liver cancer incidence has more than tripled since 1980 and is one of the most common causes of cancer death in men and women every year \cite{bray2018global}.
Computed tomography (CT) is the most commonly used modality for liver lesion detection, diagnosis, and staging. Focal liver lesions are divided into different types including malignant and benign lesions, with considerable variations in size, shape, contrast, and location. Correct discrimination between the different types of lesions is of high importance.
Manual segmentation and classification of liver lesions from CT images is very time-consuming and prone to confusion, in particular between metastasis and hemangioma lesions, which makes it a complex task.
Therefore, there is a great need and interest in automated tools to assist radiologists in the diagnosis of liver lesions from CT scans.
Recent studies in automatic liver analysis include the tasks of liver segmentation, lesion detection, lesion classification and follow-up.

% Motivation to use DNN approach
Deep learning methodologies, especially Convolutional Neural Networks (CNNs), are the top performers in most medical image processing tasks in recent years \cite{litjens2017survey}. 
% Deep learning is a data driven approach that is most effective when applied to a large training dataset. 
One of the main challenges in the medical imaging domain is the lack of sufficient amounts of annotated data for each individual medical task and application area. 
% Thanks for some recently published, publicly available datasets it was possible to encourage researchers to develop automatic tools for various medical tasks and achieve substantial progress in those fields.
Recently, several datasets in the medical field became publicly available as part of different challenges and competitions.
In the liver lesions analysis field, ISBI 2017 and MICCAI 2017 LITS challenges\footnotemark\label{fnm:1}\footnotetext{\url{https://competitions.codalab.org/competitions/15595}\label{fnt:1}.} were published and focused on the lesion segmentation task. The
%  \cite{christ2017lits} 
LiTS challenge dataset mainly consists of malignant lesions; with a goal of segmenting the lesions based on manually labeled masks provided. In the task of image classification, no open datsaset exists, thus constraining researchers to work on limited in-house datasets.

In this paper, we focus on the task of liver lesion classification using a small, in-house dataset of 332 CT slices. the goal is to categorize a given lesion into one of three types of liver lesions: metastasis, hemangioma, or cyst.

\section{BACKGROUND}

%A main challenge in Deep learning is the limited size datasets. 
Transfer learning and fine tuning are key components in using deep CNNs for medical imaging applications, for which limited size datasets exist \cite{greenspan2016guest}. In the case of insufficient data size, a common approach is to pre-train a CNN on a large dataset of natural images (such as the ImageNet dataset \cite{deng2009imagenet}, which contains millions of images from 1000 categories), and then use the trained weights as initialization following which retraining of the network may be conducted. It is possible to fine-tune all the layers of the network or freeze (keep fixed) some of them  to achieve optimized results. 

% Given two datasets from the same domain, where the first dataset is designated for a given task, and the second one is small and designated for a different task, a possible transfer learning approach would be replacing the last classification layer of the first network and retrain on the second dataset.

% Related work - finetuning on imagenet
%Fine-tuning a pre-trained network on medical data is a popular approach that has been widely used in different %medical fields as a work around the requirement of a large data set \cite{greenspan2016guest}, %\cite{litjens2017survey}.
% \cite{antony2016quantifying}, \cite{zhang2017automatic}, \cite{wang2017lung}, \cite{worrall2016automated}, %\cite{liu2016colitis} 
% Typically, the chosen model is pre-trained on a large dataset of natural images (e.g. ImageNet, PASCAL, etc.) which provides a useful starting point for the training process. For example, pre-training with ImageNet dataset was presented in \cite{bi2017automatic} for lesion segmentation and in \cite{frid2018improving} for lesion classification to improve performances.\\
% freeze weights (?)

Another challenge in the medical domain is class imbalance. In medical scans, the majority of pixels usually belong to the background class, while the informative pixels, belonging to the pathological class, are extremely under-represented \cite{greenspan2016guest}. Treating the data in a uniform manner can easily lead to overfitting. A cascaded architecture that iteratively refines the results is an effective approach to cope with this problem. For example, training one network to locate a desired ROI, followed by a second network that receives the output of the former network as input can help to simplify each network's task and make it more accurate.
% Related work - cascade architecture
Previous works have dealt with these challenges to improve liver lesions segmentation results using a cascaded learning strategy \cite{ bi2017automatic,christ2017automatic,li2017h,yuan2017hierarchical}.
% Christ et al. \cite{christ2016automatic}, 
% \cite{christ2017automatic} trained a cascaded FCN to segment the liver as ROI input for a second FCN that segment lesions. In \cite{yuan2017hierarchical}, two networks were trained to obtain a coarse and fine liver segmentation, followed by a third network that applies lesion segmentation. 
% \cite{li2017h} and \cite{bi2017automatic} also adopted a cascaded learning strategy in which a coarse liver segmentation is done as a first step prior to lesion segmentation.

In this work, we use three main steps to boost the classification performance of the network:
(1) A 2-stage cascaded architecture, in which a first network is trained to obtain liver segmentation and extract an ROI crop that is the input of a second network, that performs lesion segmentation and classification;
(2) A transfer learning framework, in which we pre-train the network on the large dataset of the LiTS challenge to produce high-quality lesion segmentation in a pixel-wise approach. We then replace the last classification layer to enable multi-class classification into different lesion categories, and train the network on the small multi-class dataset (containing three types of lesion categories), leveraging the previously learned features to fine-tune the network;
(3) A hierarchical fine-tuning pipeline using a Unique freezing protocol. We explore different strategies to optimize the fine-tuning process by freezing different parts of the network, and conclude which strategy is superior.\\
% Contributions
% Think again about the contribution of this paper compared to other papers, what is the main difference that makes it interesting
We introduce the following contributions:
\begin{itemize}
    \item We utilize a large reservoir of data aimed for lesion segmentation to improve classification results. We pre-train the network to obtain high-quality segmentation results, where the same architecture and weights are used to obtain improved lesion segmentation and classification, on a much smaller dataset. The pre-training is done on data from the same domain as the target domain (CT scans). This enables us to learn more representative features as opposed to the common approach, where pre-training is done on a large dataset from a different domain, such as ImageNet dataset (natural images).
    \item We propose a novel approach to improve segmentation and classification results by hierarchically freezing blocks of layers during fine-tuning.
\end{itemize}

\section{METHODS}

\subsection{Dataset}

We used two datasets in our study:

\textbf{1. LiTS dataset}- including 130 contrast-enhanced 3-D abdominal CT scans from the 2017 LiTS training dataset coming from several different clinical sites. The CT scans are provided with reference mask annotations (ground truth) of the liver and lesions done by expert radiologists. Additional 70 scans are provided for testing. The dataset contains $\sim$60,000 slices in total, with in-plane resolution ranges from 0.5 to 1.0 mm and slice thickness ranges from 0.7 to 5.0 mm. We further split the training dataset into 115 and 15 scans for training and validation, respectively.

\textbf{2. Sheba dataset}- including 332 2-D CT slices from the Sheba Medical Center with medical records from 140 patients for cases of cysts, metastases, and hemangiomas. Slices from healthy subjects were taken as well to be used as false examples in the training process. 
The dataset was divided as follows: 75 cysts, 71 hemangiomas, 93 metastases, 93 healthy. 
Mask annotations of liver and lesions were conducted by an expert radiologist. Different CT scanners were used with 0.71-1.17 mm pixel spacing and 1.25-5 mm slice thickness. Since the dataset size is significantly smaller than the LiTS dataset, we train and evaluate the classification results with 3-fold cross-validation.
% We split the data into 222 and 110 slices for training and validation as shown in Table 1. Fig. %\ref{fig:1} shows an example of the data used in this work.

% \begin{table}
% \centering
% \label{table:1}
% \caption{classes division into training and testing sets.}
%  \begin{tabular}{|c||c|c|c|c|c|} 
%  \hline
%   & Cyst & Hemangioma & Metastasis & Healthy\\ %[0.5ex] 
%  \hline
% %  Training & 50 & 48 & 62 & 62 \\ 
% %  \hline
% %  Testing & 25 & 23 & 31 & 31 \\
% %  \hline
% %  Total & 75 & 71 & 93 & 93 \\
%  Images & 75 & 71 & 93 & 93 \\
%   \hline
%  Patients &  &  &  &  \\
%  \hline
% \end{tabular}
% \end{table}

% \begin{figure}
% \includegraphics[scale=0.6]{data_exp} 
% \centering
% \caption{Illustration of labeled lesions data. In yellow- cyst; In blue- hemangioma; In purple- metastasis}
% \label{fig:1}
% \end{figure}

The Sheba dataset is small, yet similar to the LiTS dataset. Therefore, we expect higher-level features in the CNN to be relevant to this dataset as well. Hence, fine-tuning a pre-trained lesion segmentation network on the task of classification could significantly improve classification results with the right choice of model architecture, freezing strategy and parameters.

% \subsection{Pre-processing}
% We set the Hounsfield Unit (HU) value range to [-160, 240] according to \cite{bi2017automatic} to eliminate irrelevant information. We also apply a global standardization with the mean and std of the liver intensity values calculated over the entire dataset.

\begin{figure}[t]
\includegraphics[scale=0.65]{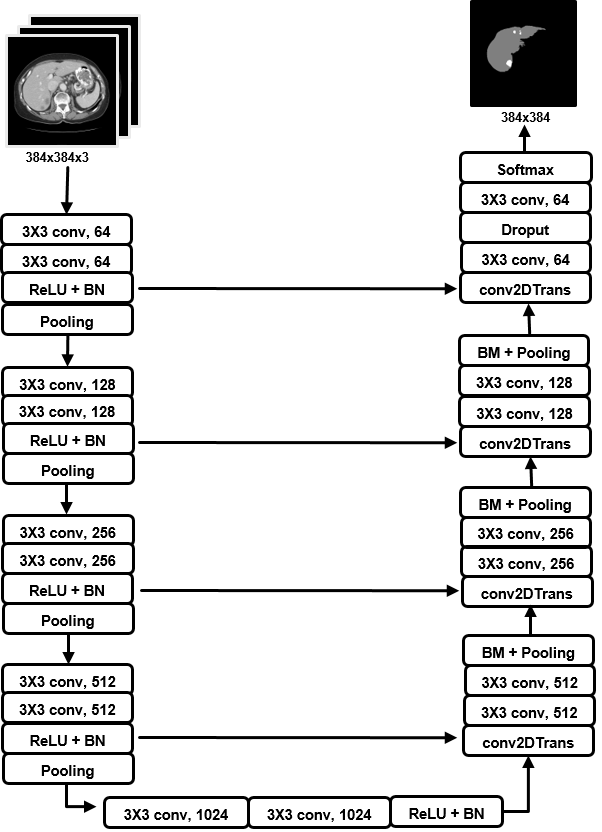}
\centering
% \balance
\caption{The U-net model used in this work for segmentation and classification.
The numbers under the images indicate the spatial dimension and the number of channels. For the convolutional layers, the kernel size and the number of filters are specified. BN stands for batch normalization.}
\label{fig:2}
\vspace{-0.3 cm}
\end{figure}

\subsection{Model Architecture}
% \vspace{-0.2 cm}
Previous works in the  medical imaging field have shown the superiority of fully convolutional networks (FCNNs) for liver lesion segmentation (e.g. \cite{ben2016fully}, \cite{christ2017automatic}) and classification (e.g. \cite{ben2018anatomical}).
% , \cite{han2017automatic} \cite{vorontsov2018liver}
FCNNs are applied to an entire input image or volume in an efficient fashion, resulting in a pixel-wise prediction map as output in the same size of the input image. 

\begin{figure*}[h]
    \includegraphics[scale=0.6]{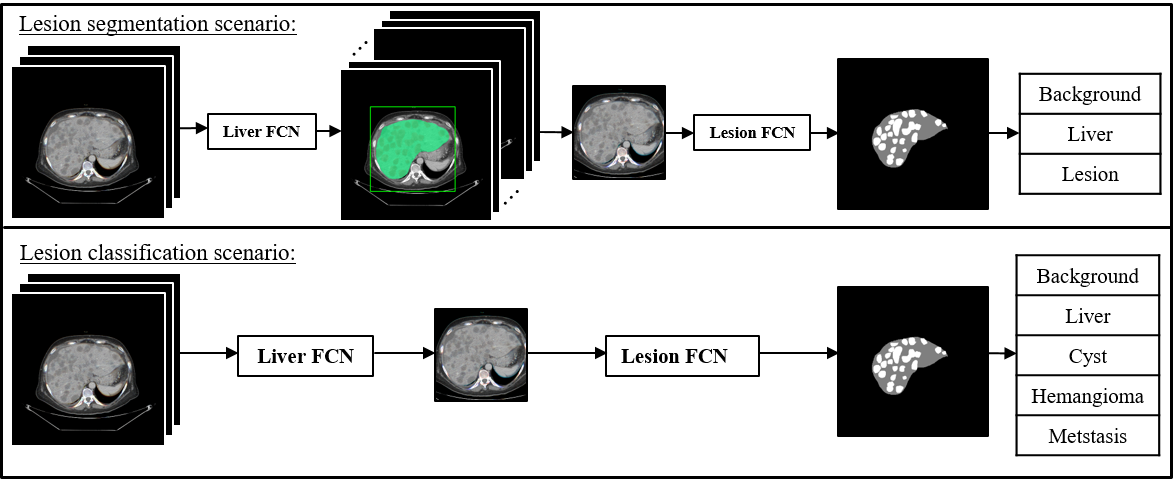}
    \centering
    \caption{Training scheme: \textbf{(1) Lesion segmentation scenario} (top)- a first FCN (Liver FCN) obtains liver segmentation per slice. Next, 2D ROIs are extracted from a 3D bounding box as input for the next FCN (Lesion FCN) that performs pixel-wise lesion segmentation; \textbf{(2) Lesion classification scenario} (bottom)- the Liver FCN obtains liver segmentation per slice and 2D ROIs are extracted and fed as inputs for the Lesion FCN that performs per-pixel lesion classification.}
    \label{fig:3}
    \vspace{-0.5 cm}
\end{figure*}

We use a U-net based model \cite{ronneberger2015u} and utilize its efficiency and simplicity for applying different experiments and manipulation. The U-net architecture is comprised of an encoder and decoder, combined with skip-connections to directly connect opposing contracting and expanding convolutional layers. The encoding part acts like a classic CNN, extracting contextual information via a hierarchy of feature maps, while the decoding part reconstructs the full image resolution via deconvolutional layers and up-sampling. This design enables the model to learn both global and local information and take fine details into account to produce quality segmentation. The full architecture is shown in Fig. \ref{fig:2}

Weighted cross-entropy loss was used as the loss function, to balance the classes, with suitable weights for each class, reversely proportional to their ratio in the dataset:
$$
L = -\sum_{i=1}^{N} {w_i^c}[\hat{P}_i^c\log{P_i^c}] \eqno{(1)}
$$
Where $P_i^c$ denotes the probability of a pixel $i$ belonging to each class $c$ and $\hat{P}_i^c$ represents the ground truth.

\subsection{Data Augmentation}
Online data augmentation was used to avoid overfitting and assist the network to be robust to lesions variability. The applied augmentations included rotations, zooming, horizontal flips, random shifting, noise addition and gray intensity modifications. The parameters used were randomly picked every epoch.

% \subsection{Post Processing}
% To further refine the results, 3D connected component was applied to the liver segmentation output of the first network.

\subsection{Implementation Details}
Due to the small size of the Sheba dataset, training was done in a 3-fold cross-validation, for better asessment of the results. A small batch size of 6 images was used for training on the LiTS dataset and a batch size of 1 was used for fine-tuning on the Sheba dataset. The number of training epochs was set to 80, with an early stopping criterion on validation improvement. Adam optimizer was used, with an initial learning rate set to 1E-4. Fine-tuning was applied with a reduced learning rate of 5E-5. Additionally, the learning rate was reduced by 10\% every 2 epochs, with a minimal threshold of 1E-8.

\textbf{Pre-processing}: We set the Hounsfield Unit (HU) value range to [-160, 240] according to \cite{bi2017automatic} to eliminate irrelevant information. We also apply a global standardization with the mean and std of the liver intensity values calculated over the entire dataset.

\textbf{Post-processing}: To further refine the results, 3-D Connected Component Analysis was applied to the liver segmentation output of the first network.

\section{EXPERIMENTS AND RESULTS}
As mentioned above, the small Sheba dataset (containing 3 lesion classes), is similar to the LiTS dataset (containing lesions deriving from one class). Therefore, we expect higher-level features in the network to be relevant to both datasets. Hence, we suggest training a strong lesion segmentation model on the large LiTS dataset and fine-tune a modified model, with the same architecture and learned weights to apply joint segmentation and classification. Since the model performs a pixel-wise classification, in which the output map represents the probability that the corresponding input pixel belongs to a certain class, the only modification we apply is expanding the output classes from 3 classes in the segmentation scenario (background/ liver/ lesion) to 5 classes (background/ liver/ cyst, hemangioma, metastasis) as illustrated in Fig. \ref{fig:3}.

In order to investigate our assumption, the experiment setup is as follows:

1. The first network was trained on the LiTS dataset to produce liver localization with high accuracy, yielding a Dice score of over 95\% on the LiTS test set.

2. The second network was trained on the liver crops generated by the first network (with the LiTS dataset as well) to obtain lesion segmentation, yielding a Dice score of over 75\% on the LiTS test set.

3. A similar model to the one in (2) was trained on the Sheba dataset (after applying liver detection using the network from (1)) to obtain lesion segmentation and classification. This model served as a baseline and its results were used as reference for followed experiments.

4. Fine-tuning was applied on the baseline model from (3), using the pre-trained model's weights from (2). Since both datasets are from the same domain, a boost in the performance was expected.

We further expanded step 4 above, by trying different freezing protocols following the weights initialization. Significant improvement of the results was observed when applying the following steps:
% \begin{itemize}
% \item Naive fine-tuning (No freezing at all).
% \item Freezing the encoder part of the network and training the decoder layers only.
% \end{itemize}
\textbf{Naive fine-tuning} - No freezing at all.
\textbf{Freeze encoder}- Freezing the encoder part of the network and training the decoder layers only.
Further improvement steps were applied for both cases by hierarchically freezing and unfreezing the weights:
By hierarchically freezing, we refer to the act of gradually freezing one block of the U-Net (a stack of 2 convolutions, ReLU, Batch Normalization, and pooling) at a time, up-to-bottom. By hierarchically unfreezing, we refer to gradually unfreezing (switching weights from fixed to trainable mode) one block at a time, bottom-up.
% \begin{itemize}
% \item \textbf{Hierarchically freezing encoder}: After one naive fine-tuning cycle, we freeze the first block of the U-net and retrain the network. We then freeze the next block and retrain again. We repeat this four times in total for the four initial blocks. 
% \item \textbf{Hierarchically unfreezing encoder}: After freezing the encoder part of the network (initial five blocks), we unfreeze the fifth block and retrain the network. We then unfreeze the fourth block and retrain, similarly to previous step and so on, until all the layers are trainable.\\
% \end{itemize}
The additional steps are the following:
\textbf{Hierarchically freezing encoder}- After one naive fine-tuning cycle, we freeze the first block of the U-net and retrain the network. We then freeze the next block and retrain again. We repeat this four times in total for the four initial blocks. 
\textbf{Hierarchically unfreezing encoder}- After freezing the encoder part of the network (initial five blocks), we unfreeze the fifth block and retrain the network. We then unfreeze the fourth block and retrain, similarly to the previous step and so on, until all the layers are trainable.\\

We compare our results with \cite{ben2018anatomical}, where a semi-supervised anatomical data augmentation framework was applied to improve classification results on the same dataset.
The evaluation was done using the same measurements as described in \cite{ben2018anatomical}: 
\textbf{Success:} number of images in which the lesion’s ground truth segmentation overlaps the model’s segmentation divided by the number of images in the test set.
\textbf{Dice1:} the average Dice segmentation measurement between lesions and not lesions, calculated per image where there is an overlap. 
\textbf{Dice2:} the average Dice segmentation measurement between lesions and not lesions, calculated per image including cases with no overlap. 
\textbf{ACC:} each image was classified based on the majority class (between the different lesion classes) and the accuracy in classification was measured.

\begin{figure}
\includegraphics[scale=0.55]{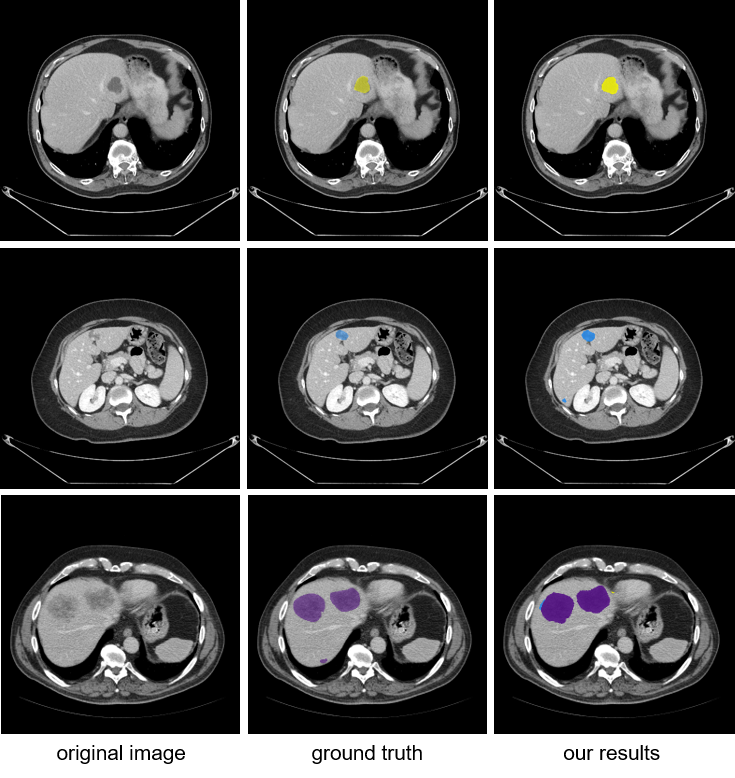} %[width=5in]
\centering
\caption{Liver lesion segmentation and classification results: \textbf{Left}: Original Image; \textbf{Middle}: Ground truth; \textbf{Right}: Our results, using hierarchical unfreezing of the encoder. \textbf{yellow}- cyst; \textbf{blue}- hemangioma; \textbf{purple}- metastasis}
\label{fig:1}
\vspace{-0.5 cm}
\end{figure}

Qualitative and quantitative results of our joint segmentation and classification method are presented in Fig. \ref{fig:1} and Table I. 
Our proposed method achieved an improvement of liver lesion segmentation and classification, compared to the baseline model training. 
Using hierarchical unfreezing of the encoder, we achieved an improvement of 15\% in Dice1, 16\% in Dice2, 5\% in the success rate, and 17\% in the classification accuracy.
Comparison of the results with \cite{ben2018anatomical} is presented in Table II. 
We show an improvement of 14\% in the success rate and 12\% in the classification accuracy.
The results show that our hierarchical fine-tuning improves segmentation and classification accuracy, with hierarchical freezing obtaining superior results. Using our method, we provided better classification results compared to an existing method. 

\begin{table}
\centering
\tabcolsep 3pt
% \resizebox{0.5\textwidth}{!}{\begin{minipage}{\textwidth}
% \label{table:2}
\caption{Performance evaluation of the different experiments}
\begin{tabular}{c c c c c c|}
\hline
\hline
& Dice1 & Dice2 & Success & Accuracy\\
\hline
Baseline & 0.49 & 0.44 & 0.89 & 0.56 \\ 
\hline
No freezing & 0.57 & 0.55 & 0.94 & 0.64 \\
\hline
Hierarchical freezing of encoder & 0.65 & 0.6 & 0.93 & 0.68 \\
\hline
Freeze encoder & 0.57 & 0.54 & 0.94 & 0.63 \\
\hline
Hierarchical unfreezing of encoder & \textbf{0.64} & \textbf{0.6} & \textbf{0.94} & \textbf{0.73} \\
% Anatomical data augmentation \cite{ben2018anatomical} & 0.83 & 0.66 & 0.8 & 0.61 \\
\hline
\hline
\end{tabular}
%\vspace{-0.8 cm}
\end{table}

%\vspace{0.1in}
\begin{table}
\centering
\tabcolsep 3pt
% \label{table:2}
\caption{Comparison of results}
\begin{tabular}{c c c c|}
\hline
\hline
 & Success & Accuracy\\
\hline
Hierarchical unfreezing of encoder  & \textbf{0.94} & \textbf{0.73} \\
\hline
Anatomical data augmentation \cite{ben2018anatomical}  & 0.8 & 0.61 \\
\hline
\hline
\end{tabular}
% \vspace{-0.4 cm}
\end{table}

\section{CONCLUSION}
In this study, we proposed a training system for joint liver lesion segmentation and classification using a large dataset for lesion segmentation (LiTS dataset) and a small dataset for lesion classification. By combining segmentation and classification with a hierarchical fine-tuning framework, we were able to improve pixel-wise classification results. 
We conclude that pre-training the network with similar data and related task, helped us learn more representative features, especially higher-level features that reside in the U-Net's encoder. 
In the future, we plan to conduct experiments to test the robustness of the scheme to additional medical tasks.

\addtolength{\textheight}{-12cm}   % This command serves to balance the column lengths

%%%%%%%%%%%%%%%%%%%%%%%%%%%%%%%%%%%%%%%%%%%%%%%%%%%%%%%%%%%%%%%%%%%%%%%%%%%%%%%%

%%%%%%%%%%%%%%%%%%%%%%%%%%%%%%%%%%%%%%%%%%%%%%%%%%%%%%%%%%%%%%%%%%%%%%%%%%%%%%%%

%%%%%%%%%%%%%%%%%%%%%%%%%%%%%%%%%%%%%%%%%%%%%%%%%%%%%%%%%%%%%%%%%%%%%%%%%%%%%%%%

\bibliographystyle{plain}
\bibliography{liverBib.bib}

\begin{thebibliography}{10}

\bibitem{ben2016fully}
A.~Ben-Cohen~et al.
\newblock Fully convolutional network for liver segmentation and lesions
  detection.
\newblock In {\em Deep Learning and Data Labeling for Medical Applications},
  pages 77--85. Springer, 2016.

\bibitem{ben2018anatomical}
A.~Ben-Cohen~et al.
\newblock Anatomical data augmentation for cnn based pixel-wise classification.
\newblock In {\em Biomedical Imaging (ISBI 2018), 2018 IEEE 15th International
  Symposium on}, pages 1096--1099. IEEE, 2018.

\bibitem{bi2017automatic}
L.~Bi~et al.
\newblock Automatic liver lesion detection using cascaded deep residual
  networks.
\newblock {\em arXiv:1704.02703}, 2017.

\bibitem{bray2018global}
F.~Bray~et al.
\newblock Global cancer statistics 2018: Globocan estimates of incidence and
  mortality worldwide for 36 cancers in 185 countries.
\newblock {\em CA: a cancer journal for clinicians}, 68(6):394--424, 2018.

\bibitem{christ2017automatic}
P.~Christ~et al.
\newblock Automatic liver and tumor segmentation of ct and mri volumes using
  cascaded fully convolutional neural networks.
\newblock {\em arXiv:1702.05970}, 2017.

\bibitem{deng2009imagenet}
J.~Deng~et al.
\newblock Imagenet: A large-scale hierarchical image database.
\newblock In {\em Computer Vision and Pattern Recognition, 2009. CVPR 2009.
  IEEE Conference on}, pages 248--255. Ieee, 2009.

\bibitem{greenspan2016guest}
H.~Greenspan~et al.
\newblock Guest editorial deep learning in medical imaging: Overview and future
  promise of an exciting new technique.
\newblock {\em IEEE Transactions on Medical Imaging}, 35(5):1153--1159, 2016.

\bibitem{li2017h}
X.~Li~et al.
\newblock H-denseunet: Hybrid densely connected unet for liver and liver tumor
  segmentation from ct volumes.
\newblock {\em arXiv:1709.07330}, 2017.

\bibitem{litjens2017survey}
G.~Litjens~et al.
\newblock A survey on deep learning in medical image analysis.
\newblock {\em Medical image analysis}, 42:60--88, 2017.

\bibitem{ronneberger2015u}
O.~Ronneberger~et al.
\newblock U-net: Convolutional networks for biomedical image segmentation.
\newblock In {\em International Conference on Medical image computing and
  computer-assisted intervention}, pages 234--241. Springer, 2015.

\bibitem{yuan2017hierarchical}
Y.~Yuan.
\newblock Hierarchical convolutional-deconvolutional neural networks for
  automatic liver and tumor segmentation.
\newblock {\em arXiv:1710.04540}, 2017.

\end{thebibliography}

\end{document}